
\magnification=1200
\hsize=6.2truein
\vsize=9.0truein
\overfullrule=0pt
\def\ref#1{$^{#1}$}

\null\vskip 1.5cm
\baselineskip=20pt
\noindent
\centerline{{\bf PHASE TRANSITIONS OF FRUSTRATED }}

\noindent
\centerline{{\bf XY SPINS IN TWO DIMENSIONS}}

\vskip 2cm



\centerline{{\it M. Benakli, M. Gabay and W.M. Saslow}\footnote\dag{{\bf Center
for
Theoretical Physics, Department of Physics, Texas A\& M University,
College Station, Texas 77843-4242} }}

\baselineskip=20pt
\parindent=60pt
\centerline{{\bf Universit\'e Paris-Sud, Centre d'Orsay}}

\parindent=60pt
\centerline{{\bf 91405 Orsay, France}}

\baselineskip=20pt

\vskip 0.4cm

The row model is used to study the commensurate-incommensurate (C-IC) and
isotropic (FFTXY) transitions of the frustrated 2D XY model on the
triangular lattice.
New relevant variables clarify the physics
of these transitions:
phase and chiral variables are coupled so that spin waves
generate long range polar interactions. The resulting dielectric constant
diverges at the transition. A single transition occurs for the FFTXY
model; in the C-IC regime the Lifshitz point is at T=0 and the C phase is
a Smectic-A like phase which disorders via a 2D nematic-smectic-A
transition.

{\bf PACS Numbers 75.10-b, 75.10.Hk}
\vfil\eject
\noindent

\noindent
 Frustrated systems have been extensively studied since they constitute
 non disordered versions of
spin-glasses[1][2].
They display
rich low-temperature phases and remarkable
phase transitions;
several non-equivalent wavevectors are
found relevant to their physics, reflecting the competing
energy scales.
As a result, frustration modifies the
naive symmetry of the Hamiltonian. For spatial dimensions $D\geq 3$,
Kawamura found by renormalization group (RG) techniques that frustrated
$O(n)$ spin models belong to a new, ``chiral"
universality class[3]. He also indicated that commensurability effects
studied by Garel and Pfeuty[4] should not play a role near these chiral
critical points.
For $D=2$ and XY spins, phase
transitions are dominated by defects. Introducing frustration results in
additional chiral variables which generate a discrete symmetry. In the
fully frustrated case the Hamiltonian for a square lattice is believed
to possess an $O(2)\times Z_2$ symmetry[5]; for the triangular lattice (FFTXY)
 one
has the extra $C_{3V}$ symmetry associated with the permutation of the three
sublattices, whence adding the possibility of a Potts transition[6]. The
transition associated with the $O(2)$ part would be Kosterlitz-Thouless
 (K-T) -like at a temperature $T_{KT}$ and the discrete $Z_2$ part
could be broken below $T_{DS}$. There is an ongoing controversy
concerning the order in which these transition should take place. RG
calculations and some
Monte Carlo (MC) simulations suggest that $T_C=T_{KT}=T_{DS}$ but two
transitions are not ruled out[7][8].
In order to
unravel the nature of the transition(s) occuring
in the fully frustrated case, various schemes have been proposed based
on the selective breaking of certain symmetries.

\noindent
In particular, the row model is a generalization
of the FFTXY model where all the bond
strengths $J$ are multiplied by $\eta$ in the horizontal direction[9]. The
FFTXY model corresponds to $\eta=1$.
In mean field theory, at low temperature
and for $\eta<0.5$ one
gets a collinear antiferromagnetic phase (C), whereas for $\eta>0.5$
an incommensurate spiral (IC) is obtained; a second order (C)-(IC)
transition line occurs for $\eta=0.5$. It extends from $T=0$ to the
Lifshitz point (LP)[10] at $T_L=1.5J$. A MC algorithm with ``self determined
boundary conditions" was developped to study this and other (IC)
structures[11] yielding an  $\eta$ versus $T$ phase
diagram quite similar to that predicted in mean field, except that the
transition temperature on the (C)-(IC) line depends on $\eta$. This
result was puzzling in view of the fact that the Lifshitz point is at $T=0$
in 2D for the ANNNI
model, that the same property
holds true for $O(n)$ spin models with $ n>2 $ as shown by RG
analysis and because of a prediction by Garel and Doniach for the 2D XY
model [12].

\noindent
In this paper we identify a new relevant variable for frustrated systems.
Its origin is described below
at $T=0$ for simplicity. In standard notations the Hamiltonian of the row model
is
$$ {\cal H} = -\sum_{<i,j>}J_{ij}cos(\theta_i-\theta_j) \eqno(1) $$  In
the ground
state, up to a global constant phase $\theta_i = \vec Q\cdot\vec r_i$
($\vec Q$ is
the modulation and $\vec r_i$ the position of the ith lattice site). To
construct a Villain-type theory[13] for frustrated XY systems we perform
local
rotations of the axes by an amount $ -\vec Q\cdot\vec r_i$. In the local
frame, all spins are ferromagnetically aligned and their phases $\phi_i$
are equal to zero.
If we excite a spin wave in the local frame (e.g along the $x$ direction),
this distorsion modifies the chirality in
the laboratory frame and {\it local} dipole fluctuations are induced.
{\it
 Spin waves have generated dipolar fields} displaying the coupling between
phase and chiral degrees of freedom. At low temperature we can estimate the
thermal averages of the induced local dipole fluctuation:
 $<\delta p>\sim T sin(\vec Q\cdot\vec r_{ij})$
 ($\vec r_{ij}$ is the nearest neighbor vector connecting sites i and j);
the polarizability $\alpha\sim 1/T\sum_k <\delta p_i \delta p_k>\sim T$ so
that the dielectric constant $\epsilon_F\sim 1+o(T/J)$[14]. This T dependence
shows that chirality  gives the dominant contribution to the total dielectric
constant (vortices of the $\phi_i$ yield $\epsilon_V\sim 1+exp-J/T$). Since the
local dipoles are generated in the chiral state
 the effect
vanishes
for the (C) case; in the spiral case, since $\vec Q$ does not vary too
strongly
with T (see below), $ <\delta p>$ increases with T; the polar
energy favors domains parallel to the dipole orientation and opposes the
ferromagnetic tendency. When polar forces overwhelm exchange forces, long
range order is lost : this occurs when $\epsilon_F$ diverges in the
direction normal to the domains. This process -- such that polar effects drive
 the transitions in the
frustrated regime -- is a new occurence of a phase
transition determined by the balance between entropy (thermally generated local
dipoles) and energy (exchange forces)[15].
For the special case
$\eta=1$ bubble domains are expected to form[16]; owing to the isotropy,
vorticies of $\phi_i$ also unbind at the transition. For $\eta$
much less than one, stripe domains are formed when the (anisotropic)
dielectric constant diverges in the $x$ direction.
This phase, such that the stiffness in the x direction is zero while that
 in the y direction is finite,
 is equivalent to the smectic-A phase of liquid
crystals. At higher temperature a smectic-A-nematic transition
takes place via a K-T melting process[17]: in spin langage the
stiffness constant in the $y$ direction goes to zero at the
paramagnetic boundary. Predictions based on our analytical
calculations are in quantitative agreement with MC simulations and support
our picture[6][7][18][19].

The starting point of an analysis a-la-Villain[13][20] is to divide
excitations into
long wavelength and short wavelength contributions.
Vortices of the $\phi_i$ and walls of the $\vec Q$ will be
taken into account for the row model in the defect part of the partition
function; as for the long wavelength part we
define  $\theta_i = \vec Q\cdot\vec r_i +\phi_i$, where $\vec Q$ is for the
moment an arbitrary vector, and extend the variations of $\phi_i$ from
$-\pi$ to $+\pi$; with this procedure the long wavelength contribution
to the partition function of hamiltonian (1) is given by
${\cal Z}=Tr_{\phi_i}exp-\beta {\cal H}_{eff}$ where
$${\cal H}_{eff}=  -\sum_{<i,j>}J_{ij}cos(\vec Q\cdot\vec r_{ij})cos(\phi_i
-\phi_j)-T Log[cosh(\sum_{<i,j>}\beta J_{ij}sin(\vec Q\cdot\vec r_{ij})
sin(\phi_i-\phi_j))] \eqno(2)$$
The first term is the contribution of the phase variables and the
second term is the new relevant (chiral) polar contribution. The Villain form
is
obtained from (2) by seeking the best hamiltonian quadratic in $\phi_i
-\phi_j$
: ${\cal H}_0=\sum_{<i,j>}\widetilde J_{ij}(\phi_i -\phi_j)^2$ [21]. The
corresponding free energy ${\cal F}$ is a function of the parameters $\vec
Q$ and $\widetilde J_{ij}$. All computational details are to appear in a
forthcoming publication[19]. The effect of the second term of (2)
is seen on Figure 1 which shows
$\widetilde J_{ij}$ as a function of distance and also its sign,
for various $\eta$. As advertized it
is a long range interaction and for large values of $r_{ij}$, $\widetilde
J_{ij}\sim 1/r_{ij}^6$ ; this feature causes the stiffness of the $\phi_i$
-- given
by $\gamma_{\vec e}=\sum_{j}\widetilde J_{ij}(\vec r_{ij}\cdot\vec e)^2$
in a direction denoted by $\vec e$ -- to be dramatically depressed.

 This is
most easily seen for the special case $\eta=1$ where, had we neglected the
(chiral)
polar contribution of (2), we would have found the incorrect low temperature
result $\gamma=\gamma_0(1-T/3J)$. Including the long range effects yields
$\gamma=\gamma_0(1-T/J[17/24-21/(16\pi\sqrt3)])$, in excellent agreement
with
MC data and with Minnhagen's prediction[6][11][18].
 As expected, the difference between
the incorrect result and the correct one is due to $\epsilon_F$ which
contributes to order T.
At $T_c\simeq 0.51J$ -- a value which agrees well with MC estimates  except
Ref.8 -- both $\gamma$ and
$\partial ^2{\cal F}/\partial Q^2$ vanish;
since $\partial ^2{\cal F}/\partial Q^2$ is proportional to the inverse
dielectric constant of the system averaged over the spin wave ensemble
(an important feature which we emphasize
 below),
the dielectric constant diverges
at $T_c$ and vortex excitations of $\phi$ and defects of $\vec Q$ occur.
Figure 2a shows the stiffness
as a function of temperature;
on the same plot we have represented the points obtained by
 performing a MC simulation of the FFTXY. The agreement extends all
the way into the critical region. We attribute this property to the fact that
$Q$ remains pinned to its $T=0$ value so that large phase fluctuations do not
occur except right at $ T_c$. The value of $T_c$ that comes out of
our equations could have been obtained by equating a dipolar
energy to the bond energy:
$\pi\widetilde J
sin(\vec Q\cdot\vec r_{ij})^2(T/3\widetilde J)^2\sim\widetilde J$.
 At $T_c$ the specific
 heat diverges; the scaling will be presented elsewhere[19].

We now turn to
the case when a (IC)-(C) transition may occur. Figure 2b shows the
stiffnesses
in the x and y directions. We notice that $\gamma_x$ goes to zero at some
temperature but the corresponding $\gamma_y$ remains finite.
As one approaches the boundary where the
dielectric constant diverges in the x direction, fluctuations in $Q_x$ become
large. Thus our curve
is not as close to the MC data near $T_c\simeq 0.22J$ as before[11][19]. On
the other hand the wavevector
remains commensurate in the y direction at all temperature. We show $Q_x(T)$
in the inset of Figure 2b;
note that it does not
go to $\pi$ at the transition. As explained in the introduction,
the homogeneous state is unstable because of polar effects. This has two
important implications. First, the dipolar fluctuation, which is proportional
to
 $sin(\vec Q\cdot\vec r_{ij})$ does not vanish and is always a relevant
 variable at higher temperature; this is why polar forces dominate. Second,
we see that because $Q_x=\pi$ is not an allowed solution, there is {\it
no incommensurate-commensurate boundary} except at $ T=0$.
 Increasing the
temperature beyond $T_c$, the
stiffness is zero in the x direction and non zero in the y direction. Along
x, polar interactions have broken the samples into stripes parallel
to y[22-25]; MC simulations in fact do show that behavior, seen in
Figure 3.
The local $Q_x$ is pinned to its Tc value, but globally there is no
rigidity along x. We can describe this situation within our formalism by
incorporating a spatial dependence to $Q_x$. The theory then ressembles
that of the dipolar magnet.
The above characteristics can be
summarized by writing the long wavelength free energy
$$F(\phi)=\int d^2r [\lambda(\partial_x\phi)^4 +
 \gamma_y(\partial_y\phi)^2] \eqno(3)$$
This is a de Gennes-like free energy of a Smectic-A liquid crystal[17]. The
 lack of
 spin order along x corresponds to the absence of translational order in the
layers.
 The layering of the smectic is the spin order in the y direction. As
shown by Day et al.[17], the smectic phase turns into a
nematic
phase above a K-T melting temperature. In our case the transition between the
``smectic-like" phase and the paramagnetic phase is in the same universality
class as the liquid crystal case. This implies that
for fixed $\eta$, as one varies the temperature, one goes from the (IC)
phase to
the ``smectic-like" phase to the paramagnetic phase. The only place where the
(IC) and (C) phases meet is at $T=0$. Thus, for the row model the Lifshitz
point is at $T=0$
for 2D XY systems[12].

 We now discuss the physical content of hamiltonian (2).
We may apply the scheme that we described to a situation where no
frustration is present, e.g for purely ferromagnetic interactions. In that
case the preferred thermodynamic $Q$ is zero, the
$\widetilde J_{ij}$ are short range, and we are simply computing the
effect of spin waves in a ferromagnet. However, as stressed above,
$\partial ^2{\cal F}/\partial Q^2$ is proportional to the inverse dielectric
constant averaged over the spin wave hamiltonian
and is always less than $\gamma$. Since in the low temperature phase
of the XY model the renormalized theory is a spin wave theory
$\partial ^2{\cal F}/\partial Q^2$ gives a fair estimate of the true value
of the dielectric constant except close to $T_c$ where vortices
contribute significantly. Applying the K-T criterion  with
$\partial ^2{\cal F}/\partial Q^2$ gives a determination of $T_c^{SW}$ that
differs from the MC $T_c$ by 14\% (for the square lattice
 we get $T_c\simeq 1.02J$ and
for the triangular lattice $T_c\simeq 1.66J$, compared to the MC
 values of 0.89J and 1.45J respectively[26]); the vortex contribution
to the dielectric constant accounts quantitatively for the difference[14].
Yet
 the spin wave hamiltonian misses the transition since neither $\gamma$ nor
$\partial ^2{\cal F}/\partial Q^2$ vanishes at  $T_c^{SW}$,
 unlike the frustrated case. The reason for the above
properties
is that we have constructed a canonical ensemble where the
macroscopic phase is free to adjust thermodynamically, as opposed to the
microcanonical procedure of fixing the phase {\it a priori}. In that
latter ensemble it is necessary to introduce a twist of the phase across the
sample or to modify the boundary conditions to extract the stiffness constants.
In the canonical ensemble these quantities appear naturally. The fluctuations
of $\vec Q$ affect both the frustrated and unfrustrated systems. In the
frustrated case these fluctuations are crucial to determine the phase
transitions. Even from the standpoint of MC simulations one sees that
"self-determined" boundary conditions are likely to produce a more effective
thermal equilibration. This is indeed what we have noticed in our MC
simulations.

To conclude, we have constructed an appropriate Villain-like theory to describe
phase transitions of modulated systems and of the fully frustrated XY model
in 2D. This is achieved by performing local rotations which align the spins
ferromagnetically in the ground state. The long wavelength fluctuations consist
of spin waves coupled to dipolar fields. These fields weaken the order by
generating an effective dielectric constant which diverges at $T_c$. Our
results
show that in the isotropic limit (FFTXY) a single transition occurs. In the
modulated case a transition is seen between a spiral phase and a smectic-A type
of order consisting of stripes of correlated spins in one direction
but without long range order in the other direction. At higher temperature a
smectic-A-nematic transition to the paramagnetic phase is expected; the
Lifshitz point is thus at $T=0$ in accord with Garel and Doniach.
In our approach the local rotations are also thermal variables. They define
the macroscopic phase of the system in the ordered state which is -- in our
derivation -- a true thermodynamic quantity. Our method is quite general and
applies to many physical situations where phase fluctuations are relevant.

{\bf Acknowledgments} We enjoyed fruitful discussions on this problem
with T. Garel who provided us with a helpful clue on the
smectic-A-nematic transition. Support from NATO grant CRG 930988 and from
IDRIS
for computer time on the Cray C98 contrat 940162 are gratefuly acknowledged.
\vfil\eject

{\bf REFERENCES}

\item{1.} J. Villain, J. Phys. {\bf C10}, 4793 (1977);{\bf C11}, 745 (1978);
J. Phys. (Paris) {\bf 38}, 26 (1977)

\item{2.} H.T. Diep, Magnetic Systems with competing interactions, World
Scientific,  in press

\item{3.} H. Kawamura, Phys. Rev. {\bf B38}, 4916 (1988);
H. Kawamura, J. Phys. Soc. Jpn. {\bf 61}, 1299 (1992)

\item{4.} T. Garel, P. Pfeuty, J. Phys. {\bf C9}, L245 (1976)

\item{5.} E. Granato, J.M. Kosterlitz and J. Poulter, Phys. Rev. {\bf B33},
4767 (1986)

\item{6.} D.H. Lee, J.D. Joanopoulos, J.W. Negele, and D.P. Landau, Phys.
Rev. {\bf B33}, 450 (1986)

\item{7.} see J. Lee, J.M. Kosterlitz, and E. Granato, Phys. Rev. {\bf B43},
11531 (1991) and references therein; B. Berge, H.T. Diep, A. Ghazali, and
 P. Lallemand, Phys. Rev. {\bf B34}, 3177 (1986)

\item{8.} G. Ramirez-Santiago, J.V. Jos\'e, Phys. Rev. {\bf B49}, 9567
(1994)
\item{9} W.-M. Zhang, W.M. Saslow, M. Gabay and M. Benakli, Phys. Rev.
{\bf B48}, 10204 (1993)

\item{10.} R.M. Hornreich, M. Luban and S. Strikman, Phys. Rev. Lett.
{\bf 35},  1678 (1975)

\item{11.} W.M. Saslow, M. Gabay and W.-M. Zhang, Phys. Rev. Lett {\bf 68},
3627 (1992)

\item{12.} W. Selke, Physics
Reports {\bf 170}, 213 (1988);
T.A. Kaplan, Phys. Rev. Lett. {\bf 44}, 760 (1980);
 T. Garel, S. Doniach, J. Phys {\bf C13}, L887 (1980)

\item{13.} J. Villain, J. Phys. (Paris) {\bf 36}, 581 (1975)

\item{14.} P. Minnhagen, Rev. Mod. Phys. {\bf 59}, 1001 (1987)

\item{15.} J.M. Kosterlitz, D.J. Thouless, J. Phys. {\bf C6}, 1181 (1973)

\item{16.} T. Garel and S. Doniach, Phys. Rev. {\bf B26}, 325 (1982)

\item{17.} A.R. Day, T.C. Lubensky, and A.J. McKane, Phys. Rev. {\bf A27},
1461
(1983); see also D.R. Nelson in Phase transitions and Critical Phenomena,
Domb and Lebowitz Vol7, p.89, Academic Press (1983)

\item{18.} P. Minnhagen, Phys. Rev. {\bf B32}, 7548 (1985)

\item{19.} M. Benakli, M. Gabay, W.M. Saslow, to be published

\item{20.} J.V. Jos\'e, L.P. Kadanoff, S. Kirkpatrick, D.R. Nelson, Phys.
Rev. {\bf B16}, 1217 (1977)

\item{21.} J. Villain, J. Phys. (Paris), {\bf 35}, 27 (1974); V.L.
Pokrovskii and G.V. Uimin, Phys. Lett. {\bf A45}, 467 (1973)

\item{22.}   M. Gabay, T. Garel, J. Phys. (Paris) {\bf 46}, 5 (1985); Phys.
Rev. {\bf B33}, 6281 (1986)

\item{23.} Y-H. Li, S. Teitel, Phys. Rev. {\bf B47}, 359 (1993)

\item{24.} A. Pimpinelli, G. Uimin and J. Villain, J. Phys.: Condens. Matter
{\bf 3}, 4693 (1991)

\item{25.} J.L. Cardy, M.P.M. den Nijs, and M. Schick, Phys. Rev. {\bf B27},
4251 (1983)

\item{26.} N. Schultka and E. Manousakis, Phys. Rev. {\bf B49}, 12071
(1994); H. Weber and P. Minnhagen, Phys. Rev. {\bf B37}, 5986 (1988);
W.Y. Shih and D. Stroud : Phys. Rev. {\bf B30}, 6774 (1984)

\vfil\eject
{\bf FIGURE CAPTIONS}
\vskip 1.0cm
FIG. 1.

$\ln (\widetilde J_{ij})$ vs $\ln (r_{ij})$ for T=0.1J; solid
line $\eta=1$, dashed line $\eta=0.575$. Insets: sign of $\widetilde
J_{ij}$ at position $\vec r_{ij}$; negative values are denoted by
circles; upper inset $\eta=1$, lower inset  $\eta=0.575$.

\vskip 1.0cm
FIG. 2.

(a) $\partial ^2{\cal F}/\partial Q^2$ vs T/J
for $\eta=1$ (solid line). Crosses: results of a MC simulation for a 30x30
FFTXY.
(b)stiffnesses along x (solid line and left scale) and along y
(dashed line and right scale)
versus T/J for $\eta=0.575$. Crosses: results of a 30x30 MC simulation. Inset
$Q_x(T)$ vs T/J.

\vskip 1.0cm
FIG. 3.

Stripe structure for $\eta=0.575$, $T=0.4$ : MC simulation of a 36x36
triangular lattice. Closed, open circles and empties denote plaquettes of
positive, negative and zero chirality respectively.
\vfil\eject

\bye